\documentclass[prb,twocolumn,floats,eqsecnum,amsmath,amssymb]{revtex4}
\usepackage[dvips]{graphicx}

\begin{document}
\title{Synchronization of delayed coupled neurons in presence of inhomogeneity}
\author{S. Sadeghi and A. Valizadeh}

\affiliation{Institute for Advanced Studies in Basic Sciences, P.O. Box 45195--1159, Zanjan, Iran}

\date{\today\vspace{1 cm}}

\begin{abstract}
In principle, while coupled limit cycle oscillators can overcome mismatch in intrinsic rates and match their frequencies, but zero phase lag synchronization is just achievable in the limit of zero mismatch, i.e., with identical oscillators. Delay in communication, on the other hand, can exert phase shift in the activity of the coupled oscillators. In this study, we address the question of how phase locked, and in particular zero phase lag synchronization, can be achieved for a heterogeneous system of two delayed coupled neurons. We have analytically studied the possibility of inphase synchronization and near inphase synchronization when the neurons are not identical or the connections are not exactly symmetric. We have shown that while any single source of inhomogeneity can violate isochronous synchrony, multiple sources of inhomogeneity can compensate for each other and maintain synchrony. Numeric studies on biologically plausible models also support the analytic results.
\end{abstract}

\maketitle
\section{INTRODUCTION}
Synchronous firing of neurons has received much attention in relation to the generation of brain wave rhythms and information processing at various aspects in the neuronal systems, such as selective attention and the binding problem\cite{brainwave1,brainwave2}. Although synchronization with nonzero phase lag has been observed between distal areas in human brains\cite{melloni}, zero phase lag synchronization is of particular interest with respect to the binding hypothesis\cite{singer}. The mechanism of these phenomena has been subject of controversial debate in a more general context; beyond its functional relevance, the zero time lag synchrony among such distant neuronal ensembles must be established by mechanisms that are able to compensate for the delays involved in the neuronal communication. Latencies in conducting nerve impulses down axonal processes can amount to delays of several tens of milliseconds between the generation of a spike in a presynaptic cell and the elicitation of a postsynaptic potential\cite{Ringo}. The question is how, despite such temporal delays, the reciprocal interactions between two brain regions can lead to the associated neural populations to fire in unison.

It is quite probable that a variety of mechanisms are responsible for bringing synchrony at different levels (distinguishing for example, among local and long-distance synchrony) and different cerebral structures. Yet there are strong evidences that long distance inter-hemispheric phase locking at zero phase lag arises via reciprocal connectivity rather than via locking to a common input \cite{Engel}. Connections between different areas and cortical columns are mediated by excitatory interactions. It has been argued that delayed excitatory connections between neuron models do not readily synchronize the neurons and in fact they usually lead to antiphase firing\cite{vrees,ernst}. This belongs to the mechanism by which model cortical neurons make the transition from excitable state to repetitive firing. Phase resetting curves (PRCs) keep track of how much an input advances or delays the next spike in an oscillatory neuron depending upon where in the cycle the input is applied. PRCs are formally found by perturbing the oscillation with a brief depolarizing stimulus at different times in its cycle and measuring the resulting phase shift from the unperturbed system\cite{Ermen96, canavier06,Izhi}. There was an agreement that the type of bifurcation which results in repetitive firing of the neuron, determines the response of the neuron to the external stimulations, so the functional form of PRC: In type I neurons with mainly positive PRC, transition to repetitive firing occurs through a saddle-node bifurcation on invariant circle (SNIC) and for type II neurons which have PRCs with a large negative lobe, repetitive firing ensues birth of a limit cycle through a Hopf bifurcation\cite{Ermen96,canavier06,Ermen01}. Just recently it is shown that type II PRCs can also occur in systems that are arbitrarily close to a SNIC bifurcation\cite{ermen2012}.

In general the form of the interaction between oscillators together with their intrinsic response (the PRC) provide sufficient information about the ability of the system to synchronize or desynchronize the oscillations\cite{nKopell,Hansel95}. On the neurons side, the type II oscillators can more easily synchronize: A PRC which contains both negative and positive lobes can allow inputs to both slow down the oscillator which is ahead and speed up the oscillator which is behind. Indeed it turned out inphase firing of type I neurons with excitatory synapses is only possible with unrealistic instantaneous couplings\cite{Tsodyks,Mirollo,Hansel95,Ermen96}. Intuitively, phase lags equal to the conduction delay are expected in causal limit when the delayed arrival of an input triggers a spike\cite{Canavier}. On the effect of the type of interaction, it is believed that inhibition and not excitation is responsible for synchronization of the neurons in many brain areas\cite{vrees,ernst}. While instantaneous inhibitory couplings lead to antiphase evolution of relaxation oscillators, in presence of delay inphase firing of is more likely to occur with inhibitory synapses\cite{vrees,ernst,Canavier,Canavier2}.

Synchronization is also affected by the configuration through which the neurons interact. A prominent example where a special configuration of the links is exploited to synchronize delayed coupled oscillators, is suggested by Fischer et al.\cite{Fischer}. In this arrangement two distant oscillators communicate indirectly through a relay oscillator and the system exhibits zero-lag synchronization between two outer elements provided the two branches are similar. While the symmetry is held (two branches are similar) the model is generic and synchronization is almost independent of the intrinsic parameters of the outer and the relay neurons and also the type of interaction. Yet, as the authors indicate the results critically depend on the symmetry of the parameters of lateral neurons and connections\cite{Fischer,relay}.

Presence of inhomogeneity in the networks of neural oscillators can also destabilize synchrony\cite{Tsodyks,Kopell98}. In classical models for synchronization coupling strength and inhomogeneity are competing factors which determine the collective dynamics of the system\cite{Kuramoto,pik}. In fact coupled oscillators can overcome finite mismatch in intrinsic frequencies and match their frequencies, but the cost is a finite phase lag\cite{pik}. How would the nonidentical neurons synchronize when they communicate from distant? We follow this question in a more general scheme by study of two model neurons coupled through delayed connections, when the parameters of either the neurons or the connections are slightly different. Using iterative maps we give a general analytic framework for the existence and stability of phase locked solutions for two phase oscillators connected by delayed pulsatile couplings, in presence of inhomogeneity. The results coincide with the those of other authors in special cases, e.g., for two similar oscillators coupled via symmetric reciprocal connections\cite{Canavier}. The numerical results given both for the phase model and for more biologically plausible conductance based models for neurons, support the the analytic findings.


\section{MODELS AND METHODS}
To model the type I and type II neurons, we have considered Wang-Buzsaki (WB)\cite{Buzsaki} and classical Hodgkin-Huxley (HH)\cite{hh} models, respectively. In type I neurons with mainly positive PRC, transition to repetitive firing occurs through a infinite period bifurcation and with a slowly increasing applied current, the neuronal dynamic changes from stationary to oscillatory with arbitrarily small frequency. For type II neurons which have PRCs with a large negative lobe, repetitive firing occurs via birth of a limit cycle through a Hopf bifurcation, and the onset of the repetitive firing occurs with non-zero frequency\cite{Ermen96} (see also Ref. [13]).

Both the WB and HH models when stimulated with a supra-threshold constant input current settle into a limit cycle, embedded in its 3- and 4-dimensional phase spaces, respectively. The current balance equation for both model neurons is
\begin{eqnarray}
C\dfrac{dv_i}{dt} &=& -g_{na}m^3 h (v_i - V_{na}) - g_k n^4 (v_i - V_k) \nonumber \\
                &&- g_l (v_i - V_l) - I_{ij} + I_{i},
\end{eqnarray}
where $i$ and $j$ are chosen from $(1,2)$. $C$ is the membrane capacitance in $\mu F/cm^2$, $v_i$ is the membrane voltage in $mV$, and $I_{i}$ is the (density of) applied current in $\mu A/cm^2$. The parameters $ g_{na} $, $ g_k $ and $ g_l$ are the maximum conductances per surface unit for the sodium, potassium and leak currents and $ V_{na}$, $ V_k$ and $V_l $ are the corresponding reversal potentials. All the constant parameters for both neuronal models are given in TABLE. \ref{para}. Details of the two models can be found in \emph{Appendix}.

\begin{table}[ht!]\footnotesize
\caption{The parameters for Wang-Buzsaki and Hudgkin-Huxley neurons}
\begin{tabular}{lcr}
    \qquad \qquad \qquad & \qquad \quad WB \qquad \qquad & \qquad \qquad HH \\
\hline
 $V_{na}$ & 55 & 55 \\
 $V_k$ & -90 & -72 \\
 $V_l$ & -65  & -50.6\\
 $g_{na}$ & 35 & 120\\
 $g_k$ & 9 & 36 \\
 $g_l$ & 0.1 & 0.3\\
 $C$ & 1 & 1 \\
 $\phi$ & 5 & - \\
\end{tabular}
\label{para}
\end{table}

The neurons are assumed to communicate through chemical synapses modeled by
\begin{equation}
\label{22}
I_{ij}= \overline{g}_{ij} s_{ij}(t - \tau_{ij}) (v_{i} - E_{syn}),
\end{equation}
where $\overline{g}_{ij}$ is the synaptic maximum conductivity and $\tau_{ij}$ is the delay in communication of the neurons $j$ (presynaptic) and $i$ (postsynaptic). Synaptic reversal potential $E_{syn}$ determines the excitatory/inhibitory type of the synapse. Throughout this manuscript, we take $E_{syn}=10 mV$ for excitatory and $E_{syn}-70mV$ for inhibitory synapses, respectively. The rate of change of the synaptic variable $s(t)$ is given by the following equation
\begin{align}
\dfrac{ds_{ij}}{dt} = \alpha f(v_j - v_{th}) (1 - s_{ij}) - \beta s_{ij},
\end{align}
with $\alpha$ and $\beta$ defining the synaptic activation and deactivation time constants, respectively, and $f(x) = 1/2 [1 + tanh(\eta x)]$ ensures activation of the synapse when the presynaptic voltage exceeds $v_{th}$. We assume the synapses are fast, i.e., the time constants of the synapses are taken much less than period of the spiking neurons\cite{hashemi}. In this regime the coupling terms can be approximated by pulsatile currents (see below).

Given the choice of model parameters in TABLE. \ref{para}, an uncoupled WB neuron exhibits periodic spiking with a period of about $16 ms$ for $I_{i}=1\mu A/cm^2 $, and a HH neuron fires with a period of about $14 ms$ for $I_{i}=10\mu A/cm^2$. Throughout this paper we have taken this typical values for $I_{2}$ and then inhomogeneity is imposed by $I_{1} = I_{2} + \Delta I$ with $\Delta I >0$. consequently, $\omega_1=\omega_2+\Delta \omega$ and the first neuron has a larger natural firing rate with $\Delta \omega>0$.

Dynamics of limit cycle oscillators can be approximated by the equation describing the evolution of the averaged phases. For an analytic inspection we use this approximation to quantify the evolution of the membrane voltage of two regularly firing neurons connected by pulse like couplings with Winfree type oscillators\cite{winfree}:
\begin{equation}
\label{win1}
\begin{array}{l}
  \dot{\theta}_i=\omega_i + g_{ij} Q(\theta_{i}) \sum_{n} \delta (t- t_{j}^{n} - \tau_{ij}),
\end{array}
\end{equation}
where $\delta$ is the Dirac's delta function describing the pulsatile interaction, and $g_{ij}$ is the strength of the synapse which can take both positive and negative values to describe excitatory and inhibitory synapses, respectively. The phase reset curve $Q(\theta)$ determines the response of the neuron to the incoming pulse. Delay time between the firing of the neuron $j$ and elicitation of a postsynaptic pulse in neuron $i$ is denoted by $\tau_{ij}$. $\omega_{i}$ is the natural firing rate of the neuron (when isolated).

We have mainly studied the synchronization of the neurons by recording intervals between spiking of two neurons. Furthermore we have measured synchrony by calculating the zero lag cross-correlation between phases of two neurons:
\begin{equation}
\label{cor}
\rho_{12} = \frac{\langle \theta_1(t) \theta_2(t) \rangle - \langle \theta_1(t) \rangle \langle \theta_1(t) \rangle} {\sqrt{Var(\theta_1)} \sqrt{Var(\theta_2)}} ,
\end{equation}
Where Var is the variance. For the limit cycle oscillators of WB and HH type, interval between two successive action potentials defines a complete cycle, and the phase increase during this time amounts to $2\pi$. Then we can assign a value to the phase by linear interpolation between two action potentials:
\begin{equation}
\label{phase}
\theta = \frac{2\pi}{T} (t - t_i),
\end{equation}
where $t_i $ is the time of the last spike and $T$ is the inter-spike interval\cite{Izhi}. Advantage of using the correlation function as a measure of phase locking is that it can discriminate between uncorrelated firing and antiphase locking since it assigns a zero value to the former and negative values to the later case. Order parameters of the Kuramoto type\cite{Kuramoto} used by other authors (see e.g. [20]) take zero value for both the antiphase and uncorrelated firings.

The differential equations were solved using a fourth-order Runge-Kutta method\cite{ode} with the step size $dt = 0.05$ of real time. Each simulation typically lasts $200 000$ iterations $ (10 sec)$.

\begin{figure}[h!]
\centering
\includegraphics[width=0.45\textwidth]{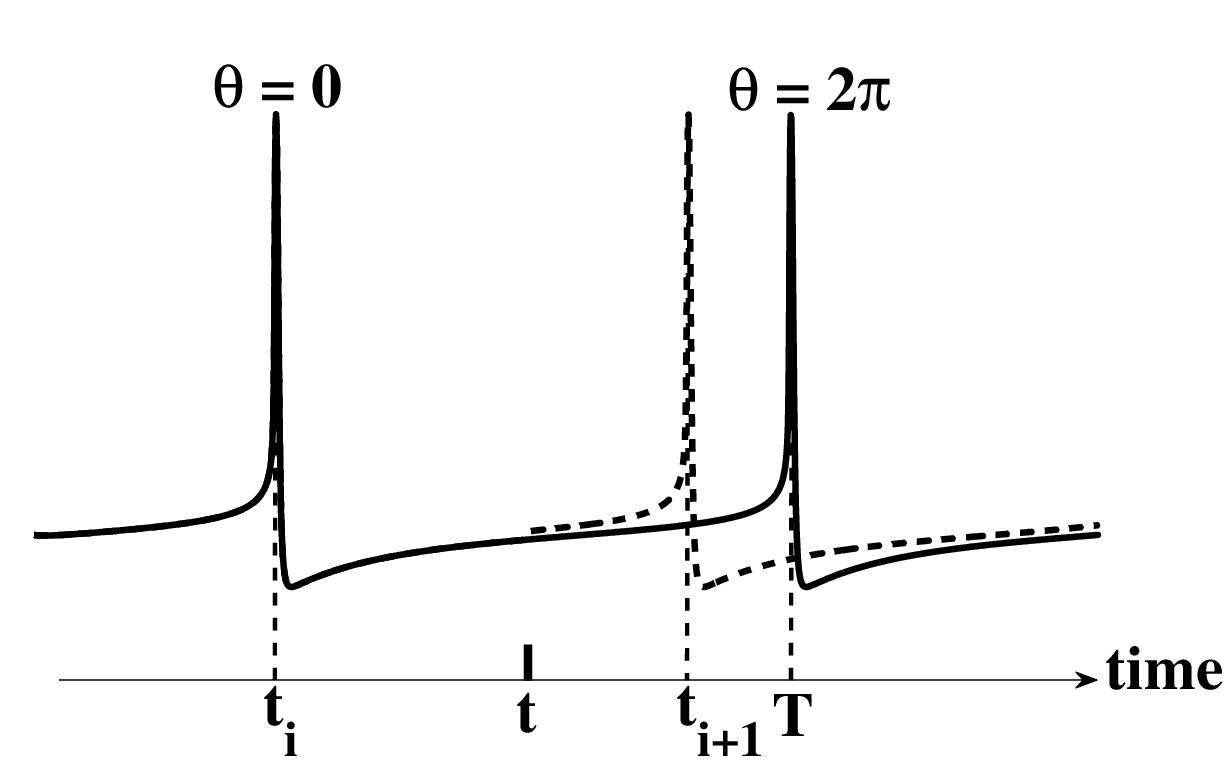}
\vspace{0cm}
\caption{Illustration of the measurement of the phase response curve of the model neuron. The solid curve indicates the voltage of the unperturbed neuron, while the dashed curve indicates the voltage after a brief perturbation is applied at time $t$.}
\label{prc}
\end{figure}

\textbf{Phase reset curves:}
Phase reset curves determine the responses of neurons to brief stimulations. As illustrated in Fig. \ref{prc}, a pulse of duration $1 ms$ and strength $h$ in different phases $\theta$ is imposed after the $i$th spike, and the timing of the next spike $t_{i+1}$ is recorded. The normalized phase reset $Q(\theta)$ is defined as
\begin{equation}
Q(\theta) = \frac{1}{h} (1 - \frac{t_{i+1} - t_i}{T}),
\end{equation}
where the phase is defined by Eq. \ref{phase} and $h$ is the strength of the pulse which can be positive or negative for excitatory and inhibitory pulses, respectively.

\begin{figure}[h!]
\centering
\includegraphics[width=0.4\textwidth]{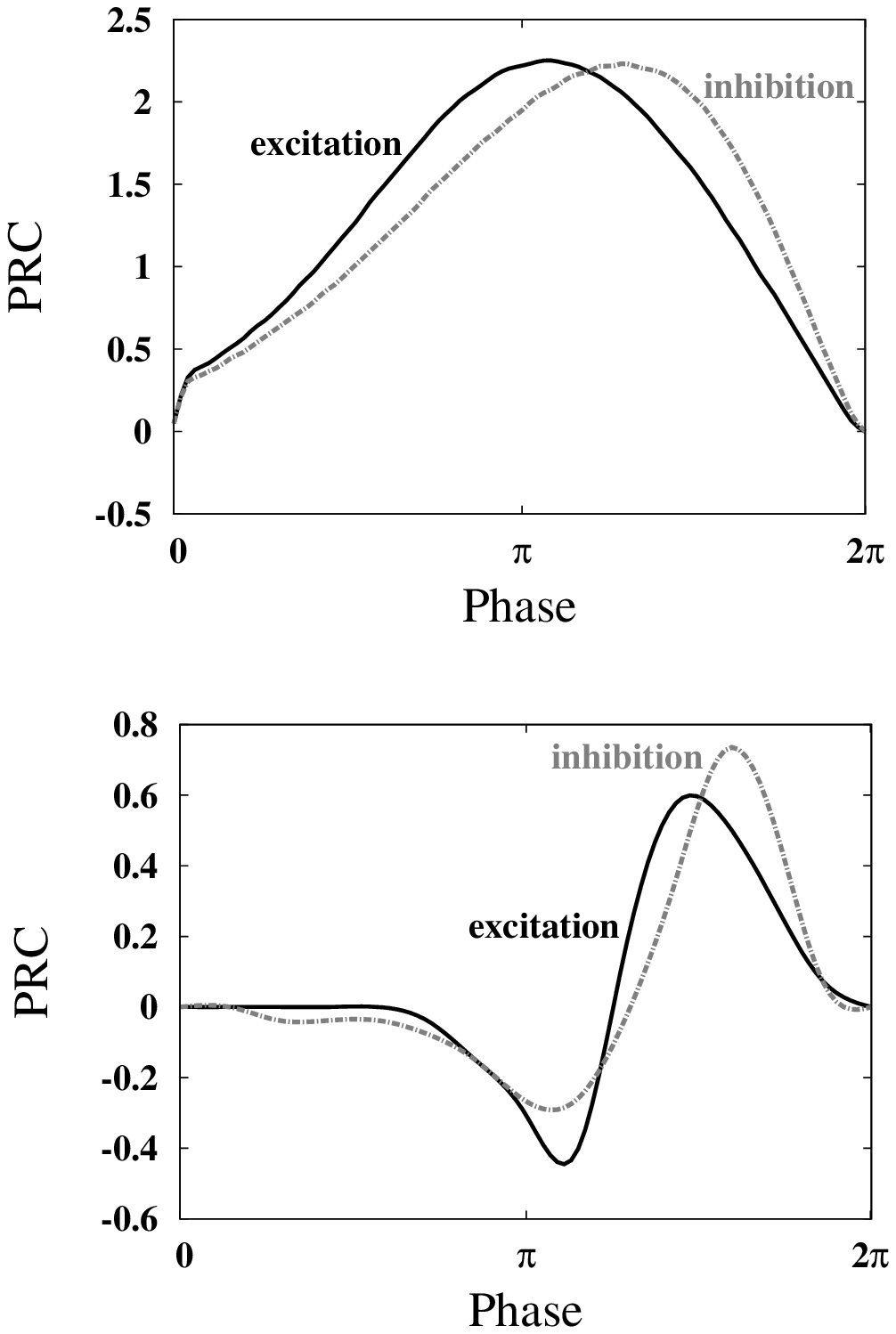}
\vspace{-0cm}
\caption{Normalized phase Reset Curves for the WB model (top) as a type-I neuron and the HH model (bottom) as a type-II neuron are plotted for excitatory and inhibitory pulses. Horizontal axis shows the phase at which a pulse of duration $1ms$ and the amplitude $0.05$ is imposed on the model neuron.}
\label{prc2}
\end{figure}

In Fig. \ref{prc2}, the phase-response curve for positive (excitatory) and negative (inhibitory) pulses are shown for the WB and HH neurons. The most notable difference between the two models is that $ PRC$ is strictly positive for the WB model, whereas it exhibits both positive and negative regions in the HH case\cite{Hansel95}. Hence, a small depolarizing perturbation always results in an advance in firing for the WB model, whereas it may either advance or delay spiking in the HH model, depending on when exactly the perturbation is delivered. Both the WB and HH models show maximal sensitivity to the external perturbation far away from the spiking events. In addition, the HH model shows more evidently that right after the spike, the system is rather unresponsive to incoming perturbations, due to refractoriness. The results for inhibitory and excitatory pulses do not show any qualitative difference.

\section{results}
\subsection{Analytic results}
\begin{figure}
\centering
\includegraphics [scale=0.4]{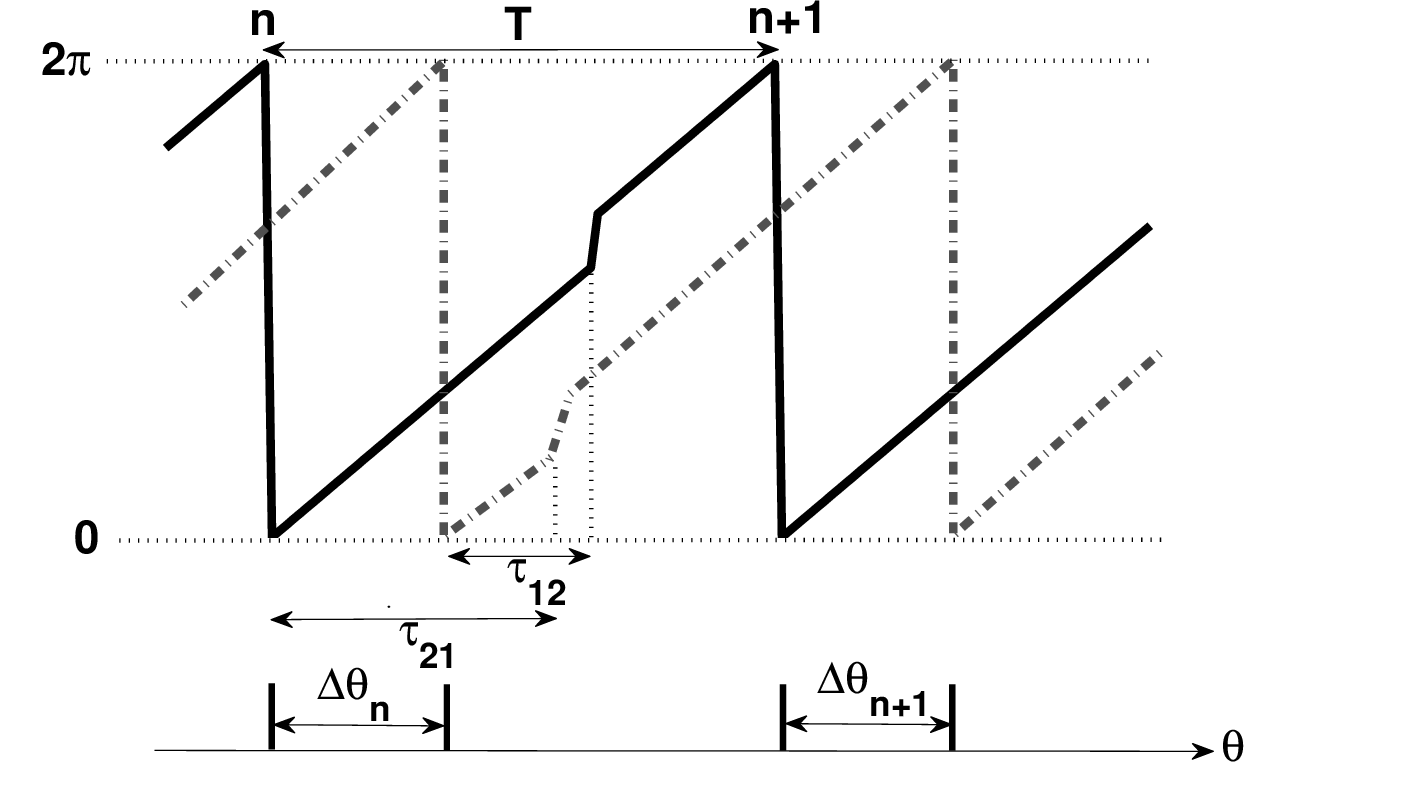}
\caption{A schematic of the map which is used for analytic investigation. The evolution of the averaged phase of two neurons is shown by solid and dashed lines, respectively. $T$ is period of firing in phase locked state and $\tau_{ij}$ is the delay time. Dynamical variable of the map is the phase difference between $n$th spikes of two neurons depicted by $\Delta \theta_{n}$.}
\label{map}
\end{figure}

Out of the causal limit (when the incoming pulses do not elicit action potentials instantaneously), we construct an iterative map by sampling the phase of the Winfree oscillators, defined in Eq. \ref{win1}, at the time of firing of the faster spiking neuron, neuron $1$. Assuming $\theta_i (t_1^n) =\theta_i(n)$ we have
\begin{equation}
\label{ap1}
\begin{array}{l}
  \theta_1 (n+1) =\theta_1 (n) + \omega_1 T + g_{12} Q(\omega_1 \tau_{12} + \Delta \theta_{n}) ,\\
   \theta_2 (n+1) = \theta_2 (n) +\omega_2 T + g_{21} Q(\omega_2 \tau_{21} - \Delta \theta_{n}).
\end{array}
\end{equation}
Here $\Delta \theta_{n}$ is the phase difference of the neurons at the instance of $n$th spike of the high frequency neuron (see Fig. \ref{map}), and $T$ is the period of the firing of the high frequency neuron which in general is different from its natural period of the firing. In the locked state $T$ is the period of the firing of the both neurons in the network (see below). 

A $1:1$ locked state is characterized by a fixed point of the map
 \begin{eqnarray} \label{ap2}
\Delta \theta_{n+1} = \Delta \theta_{n}+T \Delta \omega &+&g_{12} Q(\omega_1 \tau_{12} +\Delta \theta_{n}) \nonumber\\
&-& g_{21} Q(\omega_2 \tau_{21} - \Delta \theta_{n}),
\end{eqnarray}
which gives an implicit equation for the phase lag $\Delta \theta$ in locked state:
\begin{equation} \label{ap3}
T \hspace{.0cm} \Delta \omega = g_{21} Q(\omega_2 \tau_{21} - \Delta \theta) - g_{12} Q(\omega_1 \tau_{12} + \Delta \theta).
\end{equation}
Note that in general the network period $T$ is not constant and depends on the other parameters. Intuitively, this equation expresses that the phase difference arose from the mismatch over a period should be balanced by the mutual excitations (or inhibition) for phase locked activity. For given values of parameters, a solution of this equation, if exists, determines the phase lag of firing of two neurons in locked state. An upper limit for the mismatch which the system can tolerate in a phase locked solution is given by
\begin{equation} \label{ap4}
 T \Delta \omega < Max [g_{21} Q] - Min [g_{12} Q].
\end{equation}

\textbf{Stability condition:} Assuming a small perturbation on the phase lag in locked state $\Delta \theta_{n}=\Delta \theta + \zeta_n$, we get the linearized map
\begin{equation}
 \label{ap5}
 \begin{array}{l}
\zeta_{n+1} = \zeta_{n}[1+g_{12} Q'(\omega_1 \tau_{12} + \Delta \theta) + g_{21} Q'(\omega_2 \tau_{21} - \Delta \theta)],
\end{array}
\end{equation}
where $ Q' (\theta) = \dfrac{dQ}{d\theta}$. Stability condition is
\begin{equation}
\label{ap6}
g_{12} Q'(\omega_1 \tau_{12} + \Delta \theta)+ g_{21} Q'(\omega_2 \tau_{21} - \Delta \theta) < 0,
\end{equation}
which guaranties $|\zeta_{n+1}|<|\zeta_{n}|$.

\subsubsection{Special Cases:}

\textbf{a- Symmetric configuration:} For the bidirectional symmetric couplings $ g_{12}=g_{21}=g$ and $\tau_{12}=\tau_{21}=\tau $, and identical neurons $\omega_1=\omega_2=\omega$ Eq. \ref{ap3} gives
\begin{equation} \label{ap7}
 Q(\omega \tau+\Delta \theta)=Q(\omega \tau-\Delta \theta).
\end{equation}
This equation has always a synchronous solution $\Delta \theta=0$ regardless of the functional form of $Q$. But since $Q$ is a periodic function of phase, assuming it is continuous, other solution also exist which depends on the functional form of PRC. Stability condition for synchronous (inphase) state is $g Q'(\omega \tau) < 0$ consistent with previous results\cite{vrees,Canavier}.

\textbf{b- Symmetric couplings with dissimilar neurons:} For the bidirectional symmetric couplings $ g_{12}=g_{21}=g$ and $\tau_{12}=\tau_{21}=\tau $, Eq. \ref{ap3} gives
$$ T \Delta \omega = g [ Q(\omega_2 \tau - \Delta \theta) - Q(\omega_1 \tau  + \Delta \theta)].$$
Assuming small mismatch, we can linearize $Q$ and deduce an explicit equation for small phase lag:
\begin{equation} \label{ap8}
\Delta \theta = - \Delta \omega [\frac{\tau}{2} + \frac{T}{2g Q'(\omega \tau)}].
\end{equation}
For zero mismatch, synchronous solution exists independent of the delay and synaptic strengths. For nonzero mismatch, inphase firing is only possible if the terms in brackets cancel each other (note that second term is negative due to stability condition given below). Yet for small mismatch, near isochronous firing is possible for a range of delay times since the phase lag changes proportional to delay with the rate $\Delta \omega \ll 1$.
Stability condition given in Eq. \ref{ap6} for the symmetric connections gives:
\begin{equation} \label{ap9}
g[Q'(\omega_1 \tau + \Delta \theta)+ Q'(\omega_2 \tau - \Delta \theta)] < 0.
\end{equation}
F4or small mismatch, if a synchronous state exists, it is stable if $g Q'(\omega \tau) < 0$.

\begin{figure}
\centering
\includegraphics [scale=0.45]{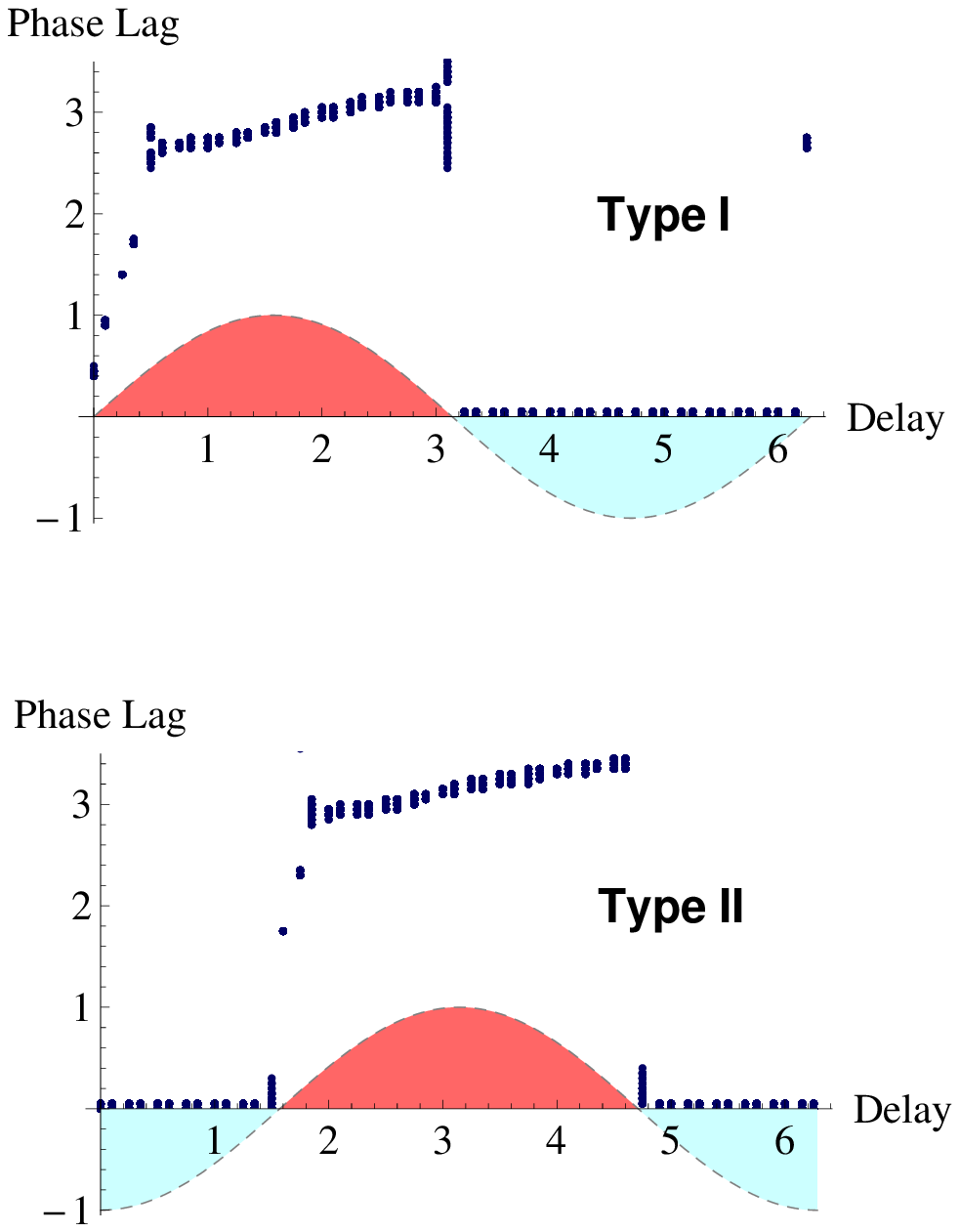}
\vspace{-0cm}
\caption{The phase difference between the spikes of two neuron is plotted against the delay in communication for type I and type II model neurons. The model neurons are identical and are of the Winfree type introduced by Eq. \ref{ap1} with $\omega_1=\omega_2=1$. The synapses are excitatory with $g_{12}=g_{21}=0.5$. The slope of phase reset curves are also shown by shaded regions, green regions show the ranges where synchronous firing is stable $gQ'(\omega \tau)<0$.}
\label{similar}
\end{figure}

\textbf{c- Near symmetric configuration:} In a near symmetric configuration when the neurons are not identical and the connections are not exactly symmetric, such that $\Delta \omega=\omega_1-\omega_2$, $\Delta g=g_{21}-g_{12}$ and $\Delta \tau= \tau_{21}-\tau_{12}$ are small parameters, the phase lag in a near inphase state can be given by a linearized approximation:
\begin{equation} \label{ap10}
\Delta \theta = \frac{1}{2 g Q'} [-(T + g \tau Q')\Delta \omega -Q \Delta g-g \omega Q' \Delta \tau],
\end{equation}
where $Q=Q(\omega \tau)$ and $Q'=Q'(\omega \tau)$. Together with the stability condition $g Q'(\omega \tau) < 0$ this equation determines deviation from synchrony for given amounts of heterogeneity in the neuronal and synaptic parameters. This equation shows that in principle it is possible to tune the phase lag by the changes in variable parameters, namely synaptic strength.

\begin{figure}
\centering
\includegraphics [scale=0.45]{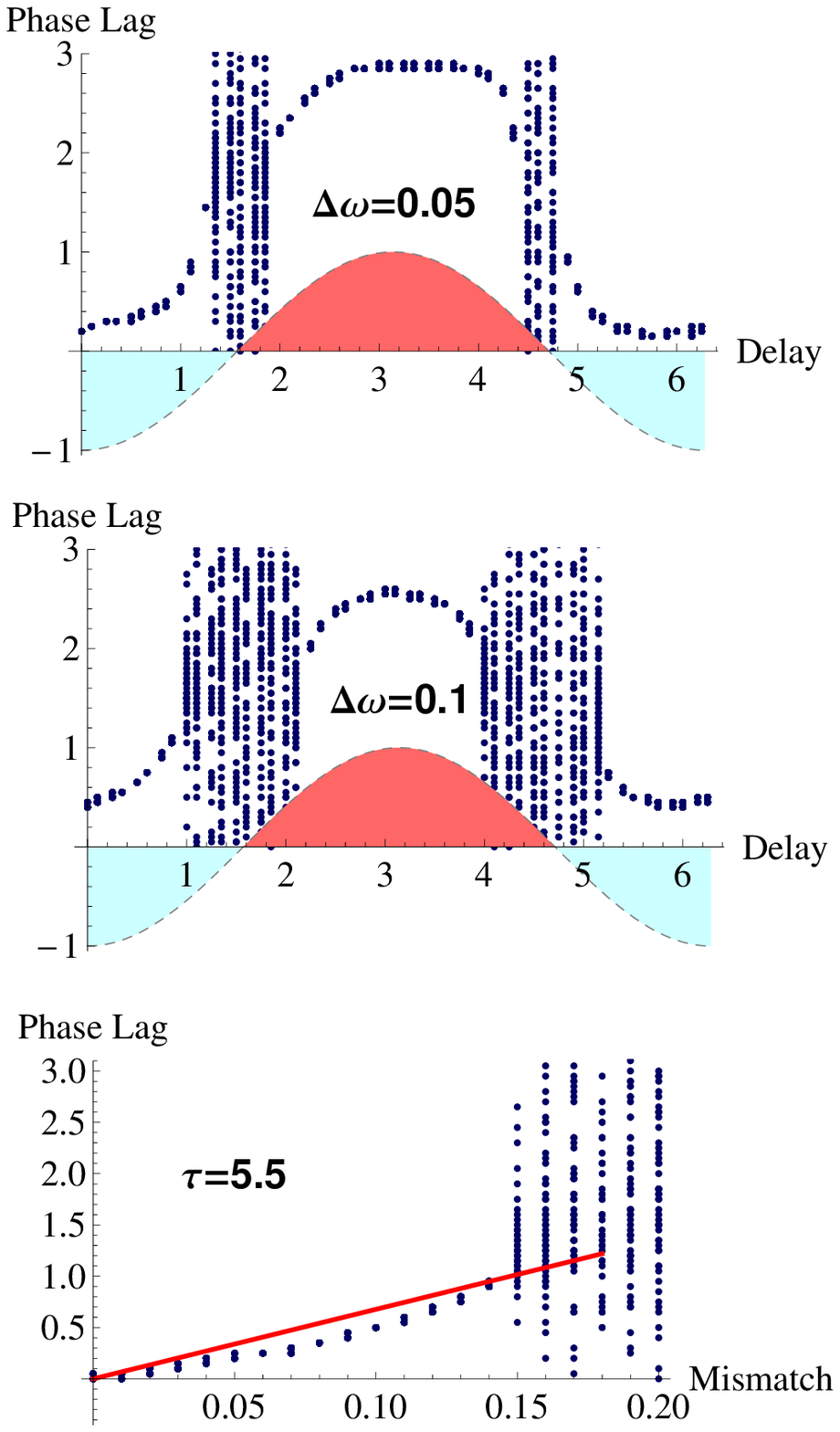}
\vspace{0cm}
\caption{Same as the Fig. \ref{similar} the phase difference between the spikes of two neuron is plotted against delay time, here for two nonidentical neurons with a mismatch shown in the plots. Lower plot shows the phase difference in respect to the mismatch for a sample value of delay time. The red solid line in the lower plot is the result of the analytic approximation Eq. \ref{ap8}.}
\label{disimilar}
\end{figure}

\textbf{d- Unidirectional coupling:} Assuming a unidirectional coupling $g_{12}=0$ necessary condition for phase locking is $T \Delta \omega < Max (g_{21} Q)$. With excitatory synapse $g_{21}>0$, this equation states that over each period, phase shift exerted by excitatory synapse on the low frequency neuron should be strong enough to compensate for the difference in natural frequency. For inhibitory synapse $g_{21} Q$ should have a positive lobe for existence of phase locked state, i.e., just type II neurons can be locked together. The phase lag in the locked state for unidirectionally coupled neurons can be explicitly found
\begin{equation} \label{ap11}
 \Delta \theta = \omega_2 \tau_{21} - Q^{-1} (\frac{ T \Delta \omega}{g_{21}}).
\end{equation}
As can be seen phase lag changes linearly with $\tau$ and if $\Delta \omega \neq 0 $, $\Delta \theta$ can be changed by $g$, i.e. in this case plasticity can tune the phase difference. On the other hand in absence of mismatch $\Delta \omega = 0 $, phase lag is independent of synaptic strength and is solely determined by firing rate and the delay time.

\subsection{Numeric results}
We first give numeric results for two pulse coupled Winfree oscillators to check the analytic results given above. Phase reset curves for type I and type II oscillators are assumed as $1-cos(\theta)$ and $sin(\theta)$, respectively. In Figs. \ref{similar} and \ref{disimilar} we have shown the time lag between the spikes of two bidirectionally coupled model neurons in symmetric and near symmetric configurations. For a fully symmetric case, when both the neurons are identical and couplings are symmetric, the results match the results of other authors\cite{Canavier}. With the excitatory synapses the neurons fire in synchrony (with zero phase lag) for the range of the delay time in which the PRC has a negative slope. This is consistent with Eq. \ref{ap3} where $\Delta \theta=0$ is a trivial solution regardless of other parameters, along with the stability criterion $gQ'(\omega \tau)<0$.

Any deviation from the symmetric case makes the phase locked state to change with other parameters as delay, synaptic constant and firing rate. For example with nonidentical neurons modeled with different firing frequencies, Eq. \ref{ap3} has not trivial solutions and the phase lag of firings in locked state changes with delay (Fig. \ref{disimilar}). For small mismatch a linear approximation can show the dependence of the phase lag on other parameters as is given in Eq. \ref{ap8}. Numeric results shown in Fig. \ref{disimilar} conform with analytic results for small mismatches. It can also be seen that transition from the inphase state to antiphase, is mediated by an unlocked region in which the neurons fire independently. This unlocked region in centered around the point in which derivative of PRC changes sign and neither inphase and antiphse solutions are stable. For the fully symmetric configuration it occurs just in the point of zero slope of PRC, but the region grows with increasing mismatch. In another point of view, the results show that the range of mismatch which the system can tolerate and remain in a phase locked state, depends on the slope of phase reset curve. Hence the tolerance of the locked state to the mismatch depends also on delay time: Near the regions in which slope of PRC changes sign, the system is vulnerable to the heterogeneity and nonidentical neurons can fire neither in synchrony nor in phase locked manner.

\begin{figure}
\centering
\includegraphics [scale=0.45]{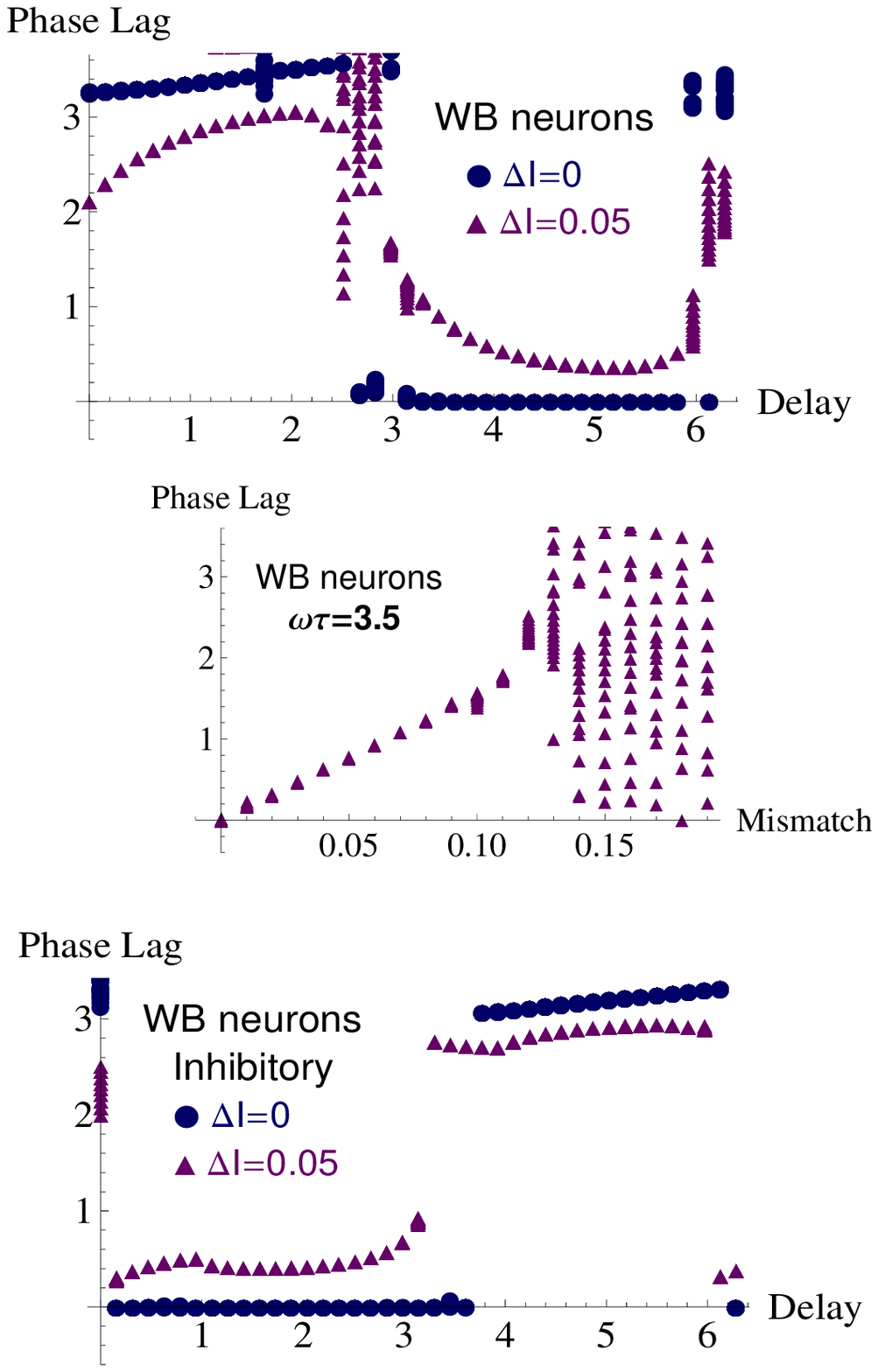}
\vspace{0cm}
\caption{Phase difference between the spikes of two neuron is plotted against delay time, for two bidirectionally coupled WB neurons with excitatory (top) and inhibitory synapses (bottom).  The results are shown for both identical neurons and in presence of mismatch. Input currents are $I_2=1$ and $I_1=1+\Delta I$. The currents are chosen such that natural period of firing for the slower neuron is about $15 ms$. Middle plot shows the phase difference vs. mismatch for WB neuron for a sample value of delay time.}
\label{wb}
\end{figure}

\begin{figure}
\centering
\includegraphics [scale=0.45]{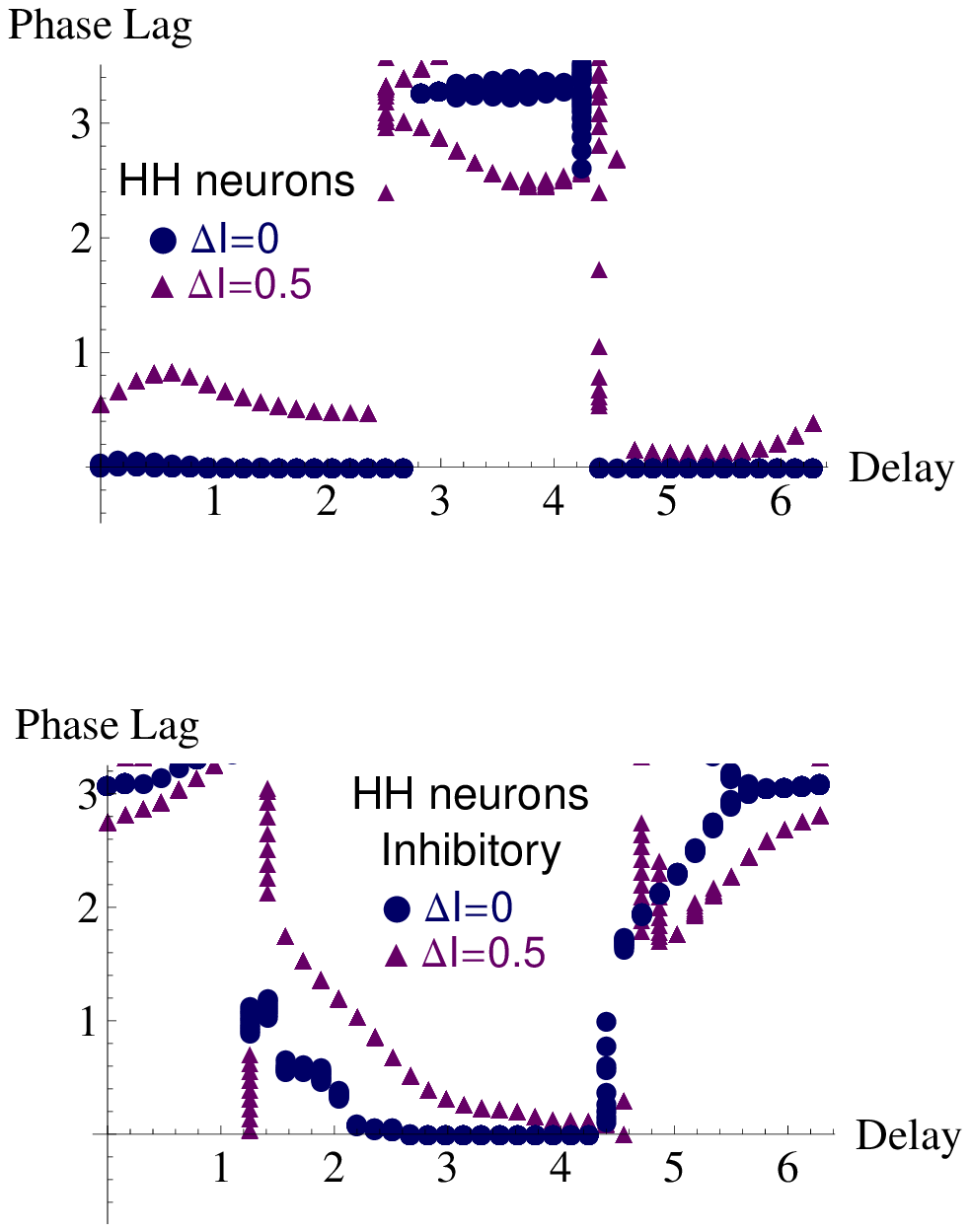}
\vspace{-0cm}
\caption{Phase difference between the spikes of two neuron is plotted against delay time, for two bidirectionally coupled HH neuron with excitatory (top) and inhibitory synapses (bottom). The results are again shown for both identical neurons and in presence of mismatch. Input currents are $I_2=10$ for WB neuron and $I_2=10$ for HH neuron and $I_1=1+\Delta I$. The currents are chosen such that natural period of firing for the slower neuron is about $15 ms$. Note that with the applied mismatch in input currents shown in this figure and Fig. \ref{wb}, relative mismatch is equal for both models $\Delta I/I=0.05$.}
\label{hh}
\end{figure}

We have repeated numeric experiments for conductance based neuronal models, WB and HH as type-I and type-II neurons, respectively. Results shown in Figs. \ref{wb} and \ref{hh} qualitatively match the results for Winfree oscillators. Synchronous state is not structurally stable and phase lag between the firings changes proportional to the mismatch for small mismatches (see the middle plot in Fig. \ref{wb}). Wang et al. have shown that with large delays synchronous firing is more stable against mismatch, comparing to small delay times\cite{Canavier2}. Our results for HH model (Fig. \ref{hh}) are consistent with this result but in a general point of view dependence of phase lag to the mismatch is determined not only by delay time but also by other parameters as coupling constant and the slope of PRC for a given delay time (see Eq. \ref{ap8}). So the form of PRC also determines the sensitivity of the synchronous state to the mismatch, for example, for a canonical form of PRC of type-II which is symmetric around $\theta= \pi$, phase lag changes similarly for small and large delays (see Fig. \ref{disimilar}).

In the formalism presented in previous section, both PRC and its derivative always appear as their multiplications with corresponding coupling constant, i.e. as $gQ$ and $gQ'$. If the PRCs maintain they form for inhibitory pulses $g<0$, with the transformation $g\rightarrow -g$ the general result for existence of phase locked solutions (Eq. \ref{ap3}) remains invariant with the simultaneous permutation transform $1\leftrightarrow2$. But the right hand side of the stability condition (Eq. \ref{ap6}) for inphase and antiphase solutions changes sign and therefore the regions of stable synchronous solutions loose stability and vice versa. For identical oscillators, for example, this means that the regions of inphase and antiphase solutions interchange. Results shown in Figs. \ref{wb} and \ref{hh} support this results: In both the figures the regions for inphase firing (near inphase firing for nonidentical neurons) are almost interchanged with regions of antiphase firing (near antiphase firing for nonidentical neurons). Specifically, with excitatory synapses and small values of delay time, typical type I neurons fire in antiphase manner while type II neurons fire synchronously.

\begin{figure}
\centering
\includegraphics [scale=0.45]{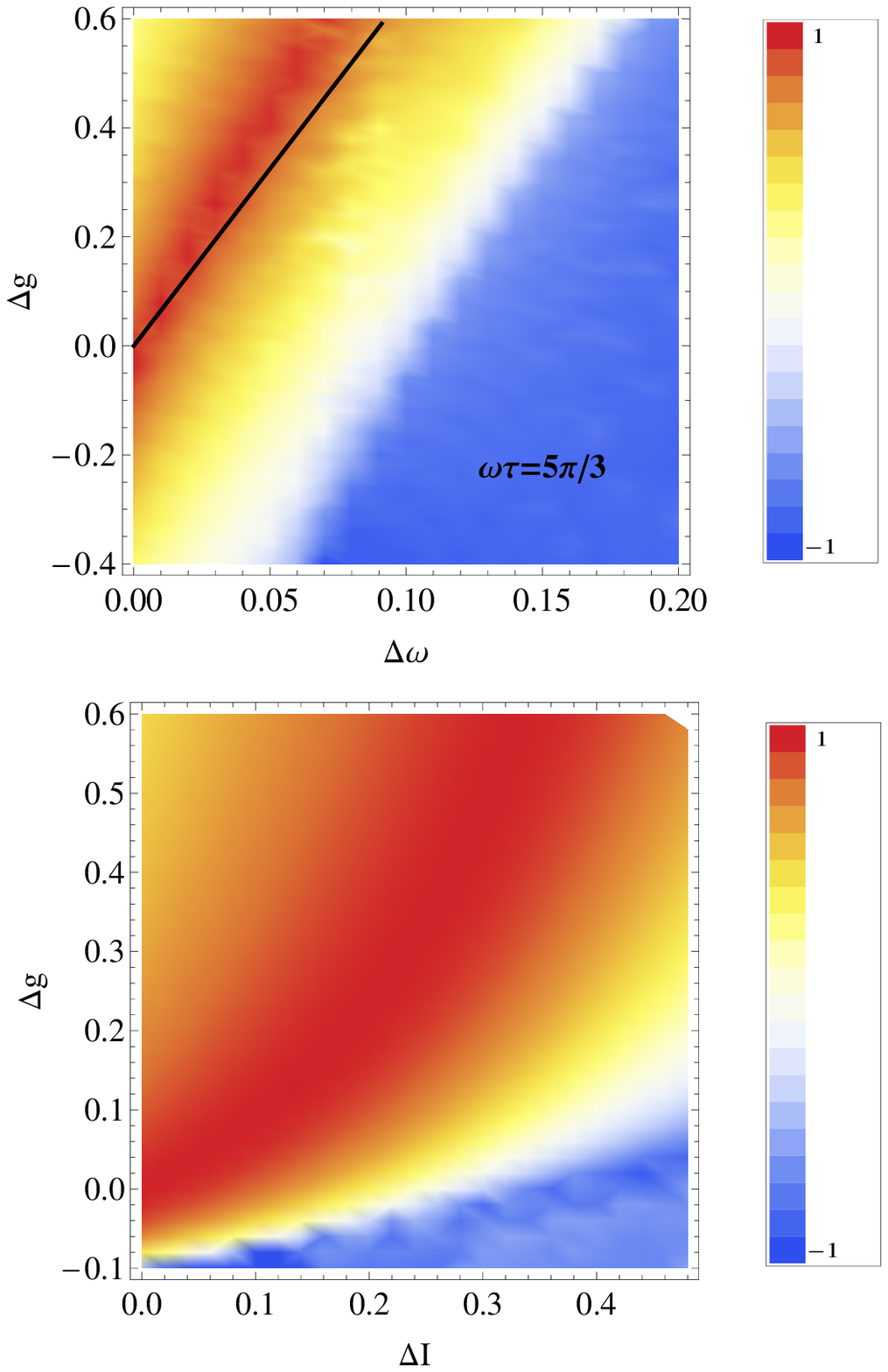}
\vspace{-0cm}
\caption{Multiple sources of inhomogeneity can compensate for each other and make synchrony. Top, A Density plot of cross correlation coefficient is shown against mismatch in firing rate and mismatch in coupling constants for two type-II Winfree oscillators. A temperature-map color scheme is used to assign warmer colors to higher values of correlation which indicate inphase synchrony. Delay is chosen such that the symmetric configuration shows inphase synchrony. It can be seen that nonidentical neurons can be synchronized by unequal coupling constants. The thick line shows result of Eq. \ref{ap10} for inphase synchrony. Bottom, the result presented for two HH neurons shows similar qualitative behavior.}
\label{density}
\end{figure}

According to Eq. \ref{ap10} it is possible to maintain synchrony, if different source of inhomogeneity show compensating effects. We have tested this hypothesis by considering nonidentical oscillators coupled by unequal strength synapses. Figure \ref{density} shows that for a given delay time if synchrony is stable for symmetric configuration, simultaneous changing of asymmetry parameters (synaptic strengths and firing rates) can maintain synchrony. Note that for the given example, the synapse from the high frequency neuron to low frequency neuron should be stronger to bring the neurons back to synchrony; an intuitively reasonable result. For the HH neurons, interestingly, the region of near inphase firing is larger comparing to Winfree oscillators but yet the results hold qualitatively.

\section{Discussion}
In most studies on the synchronization of two delayed coupled neurons, a symmetric configuration is assumed\cite{vrees,ernst,Hansel95}. This means that the neurons are assumed identical and the couplings are considered with exactly equal weights and equal delay times. To step beyond this seemingly unrealistic assumption, in this study we have considered an \emph{almost symmetric} configuration where either the neurons (nodes) or the coupling terms (edges) have slightly different parameters. Our analytic results for two pulse coupled phase oscillators show that any small inhomogeneity in a parameter, when imposed on the system, can change the synchronous firing to an \emph{almost synchronous} firing where the neurons fire with a small phase lag which is proportional to the parameter of inhomogeneity. These parameters in our minimal model are the natural firing rate of the neurons, and the strength and the delay time of two directed couplings. If for a symmetric system synchronous firing is stable, then any small asymmetric deviation in firing rates, synaptic strengths and delay times exerts a phase lag between firing of two neurons. 

As is evident from Eq. \ref{ap10}, the functional form of phase reset curve of the neurons and the strength and the type of synaptic coupling (inhibitory/excitatory) determine deviation from inphase firing. As the main outcome of this study we have shown that different sources of inhomogeneity when are present in a system together, can cancel each other and bring the system closer to synchrony. For example, for nonidentical neurons, it is recently shown (see Ref. [19]) that synchronization is only possible in the causal limit (when the excitations elicit an immediate action potential in postsynaptic neuron) with equal delays. In this case synchronization is not possible with short delay times and a symmetric increase in synaptic conductances is not a effective strategy to decrease the time lag. Instead, we have shown that an asymmetric change in synaptic strengths can bring the neurons back in synchrony (when the two first terms in the brackets in the right hand side of Eq. \ref{ap10} cancel each other). Interestingly, spike timing-dependent plasticity (STDP) exerts an asymmetric change in bidirectional synaptic couplings, but with the classical profile, STDP breaks neuronal loop and leads to the divergence of mutual couplings such that they are just restrained by limiting values\cite{bayati}. The outcome of operation of STDP in this case will always be a unidirectional coupling and hence, STDP with classical profile is not a good candidate to compensate the asymmetry in the system. It needs further investigation to check if nonlinear profiles of spike timing-dependent plasticity can show such self-regulatory effect on the inhomogeneous systems\cite{gutig}. Interestingly, axonal conduction velocity can also be modulated\cite{salami}. It is not clear if there is a self-tuning mechanism to change the conduction velocities but in principle variable delays also can role as another regulatory parameter to balance the system and bring the neurons closer to synchrony\cite{Canavier2}.

Introducing a relay neuron in the midway between two distant neuron is a sensible proposal to extend the domain of stable synchrony in the parameter space\cite{Fischer,relay}. Indeed in this case both antiphase and inphase firing of adjacent neurons, lead to simultaneous firing of wing neurons and so synchrony will be stable for almost all values of delay time. But again, in the full parameter space, this synchrony can be achieved in symmetry manifold on which all the parameters of two wings are equal. It is the subject of future study to investigate how deviation from symmetric configuration affects the synchronization properties of such a system.

\section*{Acknowledgement}
Authors gratefully acknowledge C. C. Canavier and M. McDonnell for the constructive comments and suggestions.

\appendix
\section{The model neurons}
\textbf{Wang-Buzsaki (WB) Model:} The steady-state activation $m_{\infty}$ and the rate equation for the inactivation variable $h$ in the expression for sodium current and rate equation for the activation variable $ n$ in the expression for potassium current are given respectively as follow:
\begin{equation}
\begin{array}{l}
m = m_\infty (V) = \dfrac{\alpha_m (V)}{\alpha_m (V) + \beta_m (V)},\\
\dfrac{dh}{dt} = \phi [\alpha _h (V) (1 - h) - \beta _h (V) h],\\
\dfrac{dn}{dt} = \phi [\alpha _n (V) (1 - n) - \beta _n (V) n].\\
\end{array}
\end{equation}
The rate constants for $ m_\infty $, $ h $, and $ n $ are:
\begin{equation}
\begin{array}{l}
\alpha_m (V) = -0.1 (V + 35)/(\exp ( -0.1(V + 35) ) - 1),\\
\beta_m (V) = 4 \exp (- (V +60)/18),\\
\\
\alpha_h (V) = 0.07 \exp (- (V + 58) )/20),\\
\beta_h (V) = 1/ (\exp (-0.1 (V +28) + 1),\\
\\
\alpha_n (V) = -0.01 (V + 34)/(\exp ( -0.1(V + 34) ) - 1),\\
\beta_n (V) = 0.125 \exp (- (V +44)/80).\\
\end{array}
\end{equation}

\textbf{Hodgkin-Huxley (HH) model:} The rate equations for the activation variable $ m $ and inactivation variable $ h $ of the sodium expression, and $ n $, activation variable of potassium, obey the differential equations:
\begin{equation}
\begin{array}{l}
\dfrac{dm}{dt} = [\alpha _m (V) (1 - m) - \beta _m (V) m],\\
\dfrac{dh}{dt} =  [\alpha _h (V) (1 - h) - \beta _h (V) h],\\
\dfrac{dn}{dt} =  [\alpha _n (V) (1 - n) - \beta _n (V) n].\\
\end{array}
\end{equation}
The rate constants for $ m $, $ h $, and $ n $ are:
\begin{equation}
\begin{array}{l}
\alpha _m (V) = {{(2.5 - 0.1V)}}/{{(\exp{(2.5 - 0.1V)}  - 1)}},\\
\beta _m (V) = 4\exp{ {(-V}/{{18)}}},\\
\\
\alpha _h (V) = 0.07\exp{ {(-V}/{{20)}}},\\
\beta _h (V) = {1}/{{(\exp{(3 - 0.1V)}  + 1)}},\\
\\
\alpha _n (V) ={{(0.1 - 0.01V)}}/{{(\exp{(1 - 0.1V)}  - 1)}},\\
\beta _n (V) = 0.125\exp{ {(-V}/{{80)}}}.\\
\end{array}
\end{equation}


\end{document}